\documentclass[11pt]{article}
\usepackage{graphicx}
\usepackage[margin=1.25in]{geometry}
\usepackage[usenames,dvipsnames]{color}
\usepackage{url}
\usepackage{hyperref}
\usepackage{cleveref}

\crefname{section}{Sec.}{Secs.}
\crefname{chapter}{Chapter}{Chapters}
\crefname{figure}{Fig.}{Figs.}
\crefname{equation}{Eq.}{Eqs.}
\crefname{table}{Table}{Tables}
\crefname{appendix}{Appendix}{Appendices}

\newcommand{\lsim}{\mathrel{\hbox{\rlap{\lower.75ex \hbox{$\sim$}} \kern-.3em \raise.4ex \hbox{$<$}}}}
\newcommand{\gsim}{\mathrel{\hbox{\rlap{\lower.75ex \hbox{$\sim$}} \kern-.3em \raise.4ex \hbox{$>$}}}}

\def\Title#1{\begin{center} {\LARGE #1 } \end{center}}
\def\Author#1{\begin{center}{ \sc #1} \end{center}}
\def\Address#1{\begin{center}{ \it #1} \end{center}}

\newenvironment{Abstract}{\begin{quotation} \begin{center}
                       ABSTRACT
     \end{center}\bigskip  }{\end{quotation}}

\def\beq{\begin{equation}}
\def\eeq#1{\label{#1}\end{equation}}
\def\eeqn{\end{equation}}

\newenvironment{Eqnarray}%
   {\arraycolsep 0.14em\begin{eqnarray}}{\end{eqnarray}}
\def\beqa{\begin{Eqnarray}}
\def\eeqa#1{\label{#1}\end{Eqnarray}}
\def\eeqan{\end{Eqnarray}}

\let\bar=\overbar

\def\lsim{\mathrel{\raise.3ex\hbox{$<$\kern-.75em\lower1ex\hbox{$\sim$}}}}
\def\gsim{\mathrel{\raise.3ex\hbox{$>$\kern-.75em\lower1ex\hbox{$\sim$}}}}

\def\del{\partial}
\def\Dslash{\not{\hbox{\kern-4pt $D$}}}
\def\dslash{\not{\hbox{\kern-2pt $\del$}}}
\def\pslash{\not{\hbox{\kern-2pt $p$}}}
\def\ETmiss{\not{\hbox{\kern-4pt $E$}}_T}

\def\Dlr{\mathrel{\raise1.5ex\hbox{$\leftrightarrow$\kern-1em\lower1.5ex\hbox{$D$}}}}

\def\MSB{{\bar{M \kern -2pt S}}}
\def\msb{{\bar{\scriptsize M \kern -1pt S}}}

\def\drb{{\bar{\scriptsize D \kern -1pt R}}}

\newcommand\snowmass{\begin{center}\rule[-0.2in]{\hsize}{0.01in}\\\rule{\hsize}{0.01in}\\
\vskip 0.1in Submitted to the  Proceedings of the US Community Study\\ 
on the Future of Particle Physics (Snowmass 2021)\\ 
\rule{\hsize}{0.01in}\\\rule[+0.2in]{\hsize}{0.01in} \end{center}}

\begin{document}


\Title{{Prompt electron and tau neutrinos and antineutrinos in the forward region at the LHC}}

\bigskip 

\Author{Weidong Bai$^a$, Milind Vaman Diwan$^b$, Maria Vittoria Garzelli$^c$, Yu~Seon~Jeong$^d$, Karan Kumar$^e$, Mary Hall Reno$^f$}

\medskip

\Address{$^a$Sun Yat-sen University,  School of Physics, 
 No. 135, Xingang Xi Road, Guangzhou, 510275, P. R. China\\
 $^b$Brookhaven National Laboratory, Upton, New York, USA\\
$^c$Universit\"at Hamburg, II Institut f\"ur Theoretische Physik, 
Luruper Chaussee 149, D-22761, Hamburg, Germany\\
$^d$Chung-Ang University, High Energy Physics Center, 
Dongjak-gu, Seoul 06974, Republic of Korea\\
$^e$Department of Physics and Astronomy, Stony Brook University, Stony Brook, NY 11794, USA\\
$^f$Department of Physics and Astronomy, University of Iowa, Iowa City, IA 52242, USA}

\medskip

\begin{Abstract}
\noindent 
Neutrino fluxes at high rapidity and at high energy are sensitive to QCD dynamics of heavy-flavor production in kinematic regions where measurements have not yet been made. The
FASER$\nu$ and SND@LHC 
experiments 
scheduled for Run 3 at the LHC and the proposed Forward Physics Facility with a suite of experiments 
 during the High-Luminosity LHC phase 
will probe neutrinos at high pseudorapidity. 
This short paper reports on recent evaluations of the prompt $\nu_\tau+\bar\nu_\tau$ and
$\nu_e+\bar\nu_e$ double-differential cross sections in $pp$ collisions at the Large Hadron Collider from the production and decays of $D_s^\pm$ and $D^\pm$, respectively. For $\sqrt{s}=14$ TeV, the double-differential neutrino energy and pseudorapidity distributions are evaluated at NLO QCD. Data tables with these predictions are presented. Future work needed to refine predictions of neutrino and antineutrino fluxes in the forward region at the LHC is discussed.
\end{Abstract}

\snowmass

\vfil\eject

\section{Introduction}

The far-forward region at the Large Hadron Collider (LHC) is a complementary region to the central and forward regions covered by the ATLAS, CMS and LHCb main detectors. Particle detection in the far-forward region probes fundamental QCD and physics beyond the standard model (BSM) \cite{Anchordoqui:2021ghd,Feng:2022inv}. Besides neutral hadrons and photons, as probed by LHCf, interesting messengers of far-forward inelastic-scattering physics are neutrinos~\cite{DeRujula:1984pg,DeRujula:1992sn,Park:2011gh}. In this respect, two experiments for the far-forward region at the LHC have been approved for Run~3: FASER$\nu$~\cite{Abreu:2019yak,Abreu:2020ddv} and SND@LHC~\cite{Ahdida:2020evc,Navarria:2022krz}. 
They have the potential to detect neutrinos that come from $pp$ collisions at the LHC, extracted along the direction tangent to the beam line. The prototype FASER$\nu$ experiment deployed in 2018 that collected data from 12.2 fb$^{-1}$ of $pp$ collisions at the LHC already reported evidence of some candidate neutrino interaction events \cite{FASER:2021mtu}, providing a first hint of the fact that more systematic measurements of events induced by LHC neutrinos should be actually feasible during forthcoming Runs.

Neutrinos are produced both by $W^\pm$ and $Z$ boson decays, and by light- and heavy-hadron decays. It has been shown with a {\texttt{Pythia8.2}} \cite{Sjostrand:2014zea} evaluation  that for pseudorapidities larger than $|\eta_\nu |>6.7$, heavy flavor significantly dominates over $W^\pm$ and $Z$ boson production of neutrinos \cite{Beni:2019pyp}. 
In particular, $D_s^\pm$ decays and to a lesser extent \cite{Bai:2020ukz}, $B$-meson and other $D$-meson decays, are essentially the only sources of $\nu_\tau$ and $\bar{\nu}_\tau$ for such large $|\eta_\nu|$ values. Thus, measurements of $\nu_\tau$ and $\bar{\nu}_\tau$ events will 
probe heavy-flavor production and decay in kinematic regimes in which no dedicated measurements at accelerators have been made so far.
In addition, the high energy flux of $\nu_e+\bar\nu_e$ in the very forward region comes predo\-mi\-nan\-tly from $D^\pm$, $D^0$ and $\bar D^0$ production and decays that include electron neutrinos in the final state. Towards more central rapidities, $B$-meson contributions form a larger fraction of the $\nu_e+\bar\nu_e$ flux. The high-energy flux of $\nu_e+\bar\nu_e$, is expected to be large enough to be also usable for probing heavy-flavor production and decay in the very forward region.

Measurements of neutrino fluxes in the far-forward region will yield a better understanding of forward heavy-flavor production in $pp$ collisions, useful to understand QCD dynamics and the interplay of perturbative and non-perturbative aspects, particularly as they influence the low transverse momentum and large rapidity distributions of the heavy hadrons~\cite{Bai:2020ukz}. Parton distribution functions will be probed in both small and large longitudinal momentum fraction $x$ regimes. New measurements of forward neutrino fluxes will ultimately help to better understand $\nu$-induced deep-inelastic scattering, both in inclusive and heavy-flavor production. Furthermore, detailed studies and data analyses will require and lead to improved soft physics models in shower Monte Carlo codes, as well as to new tunes of their input parameters, with consequences for many other collider studies.

Planning is underway for experiments at the proposed Forward Physics Facility \cite{FPFLong:2022tbd}. In order to develop detector design and optimize detector placement within the facility, tabulated results for neutrino distributions as functions of both the neutrino energy $E_\nu$ and neutrino pseudorapidity $\eta_\nu$ are of great utility.
Tables with state-of-the-art QCD predictions for $d^2\sigma/(dE_\nu d\eta_\nu)$ for $pp$ collisions at $\sqrt{s}=14$ TeV are presented in this work and made available to the scientific community.

\section{Ingredients of the computation of $D_s^\pm$ production}

For the neutrino fluxes presented here, we evaluate the single-particle inclusive charm quark distributions including NLO QCD radiative corrections \cite{Nason:1989zy}. 
The $pp$ scattering process is dominated by gluon-gluon fusion.
The collinear parton model treats the incoming partons as strictly collinear with the proton beam direction. When looking into the very forward direction, even a small transverse momentum deviation from the collinear approximation can have an impact on the number of events. We implement a phenomenological Gaussian smearing of the charm-quark transverse momentum  as described in Refs.~\cite{Bai:2020ukz,Bai:2021ira}. Fragmentation is implemented in the colliding parton center-of-mass frame using the Peterson fragmentation functions \cite{Peterson:1982ak}.
Details of the charmed-meson decay implementation are described in Ref. \cite{Bai:2020ukz}. Fragmentation fractions are assumed to be $B(c\to D_s^+)=0.0802$, $B(c\to D^+)=0.2404$ and $B(c\to D^0)=0.6086$ \cite{Lisovyi:2015uqa}.
We find that for a Gaussian-like smearing function of the 
charm-quark momentum, an average transverse momentum of $\langle k_T\rangle=0.7$ GeV \cite{Bai:2020ukz}, together with Peterson fragmentation of the quark to charmed meson, gives transverse momentum and rapidity distributions of the charmed mesons similar to those obtained by a NLO QCD calculation matched to parton showers according to the POWHEG approach~\cite{Frixione:2007vw} interfaced to PYTHIA8 \cite{Sjostrand:2014zea,Sjostrand:2019zhc} to simulate parton shower emissions beyond the first one, hadronization and further soft-physics effects.

Our default parton distribution functions (PDFs) are the PROSA19 NLO PDFs \cite{Zenaiev:2019ktw}. These PDFs are based on fits to a range of datasets that includes, among others, open heavy-flavor production at LHCb. The renormalization and factorization scales $\mu_R$ and $\mu_F$ used in the PROSA19 fits  is
\begin{equation}
    m_{T,2} = \sqrt{(2m_c)^2+p_T^2}\,.
\end{equation}
Our default central scale is consistently chosen as $\mu_R=\mu_F=m_{T,2}$.
This scale choice allows for an improved perturbative convergence of the predictions in the bulk of the phase-space, with respect to standard choice  $\mu_R = \mu_F = m_T$ more often considered.
The PROSA fit was performed assuming 
$\langle k_T\rangle =0$ and using ratios between differential distributions. A compa\-rison of our predictions for $D_s^\pm$ production, including NLO QCD radiative corrections, with LHCb  data for $D_s^\pm$ absolute transverse momentum and rapidity distributions suggest that a somewhat larger value of $\langle k_T\rangle$, i.e. $\langle k_T\rangle= 0.7$,  leads to an increased agreement of theory predictions with experimental data. An even larger agreement with the LHCb $D_s^\pm$ data is achieved with $\mu_F=2\mu_R=2m_T$ ($m_{T} = \sqrt{m_c^2+p_T^2}$) and $\langle k_T\rangle = 1.2$ GeV. Results for this scale and $\langle k_T\rangle$ values are also shown in the following, besides those with our preferred inputs.

\begin{figure}
    \centering
    \includegraphics[width=0.95\textwidth]{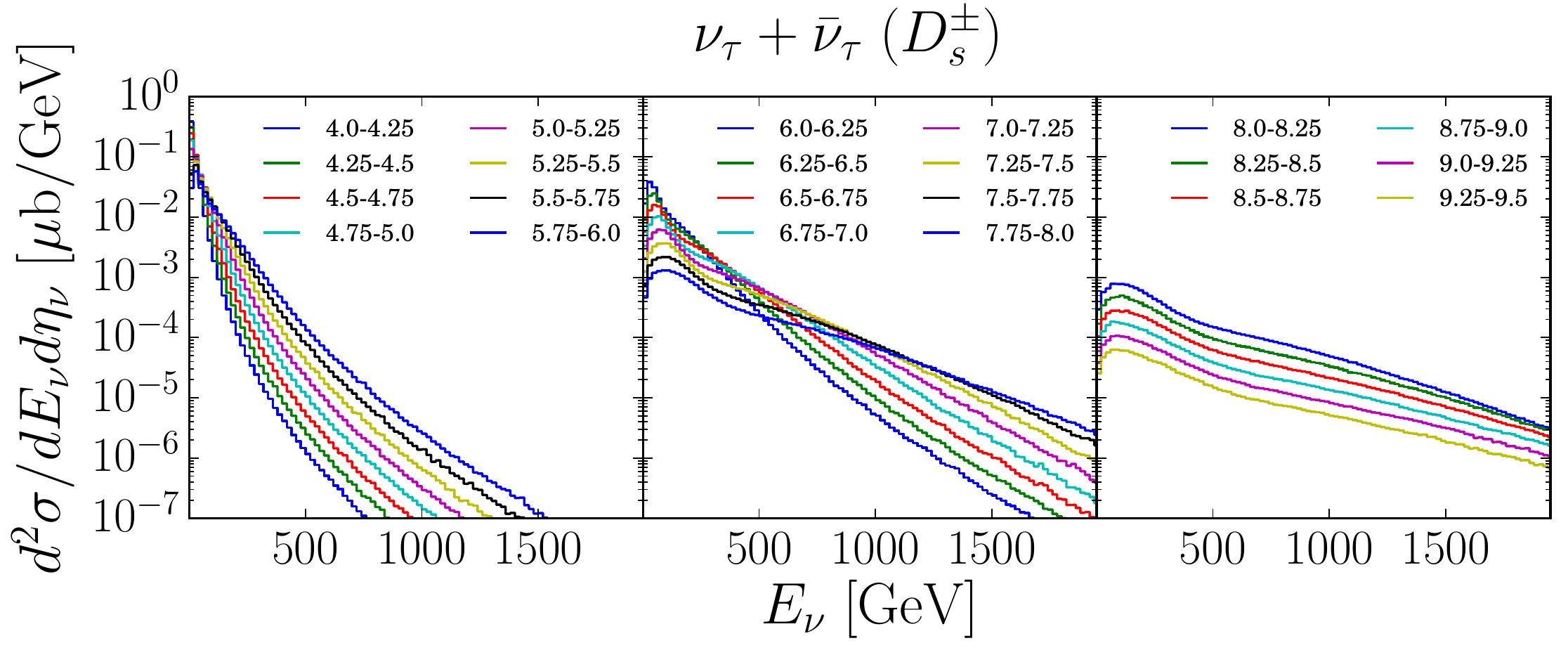}
    \includegraphics[width=0.95\textwidth]{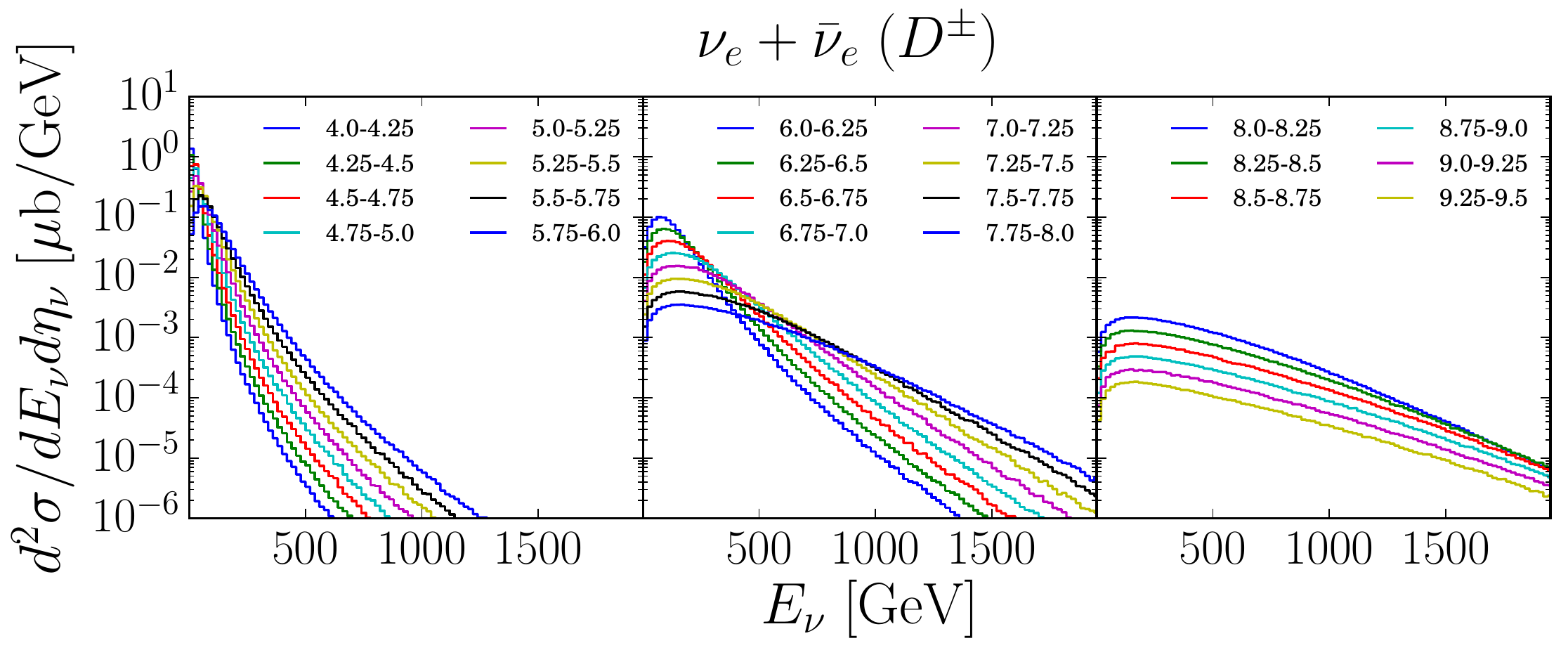}
    \caption{The double-differential distributions of the sum of $\nu_\tau+\bar\nu_\tau$ from $D_s^\pm$ (upper) and $\nu_e+\bar\nu_e$ from $D^\pm$ (lower) production and decay as a function of neutrino energy for $\eta_\nu=4.0-4.25$ through $\eta_\nu=9.25-9.5$. Here,  the default PROSA PDF scales  $\mu_R=\mu_F=m_{T,2}$ and $\langle k_T\rangle=0.7$ GeV are used for the NLO QCD evaluation for $\sqrt{s}=14$ TeV in $pp$ collisions at the LHC.}
    \label{fig:3in1}
\end{figure}

\section{Predictions}

\begin{figure}
    \centering
    \vskip 0.25in
\includegraphics[width=0.45\textwidth]{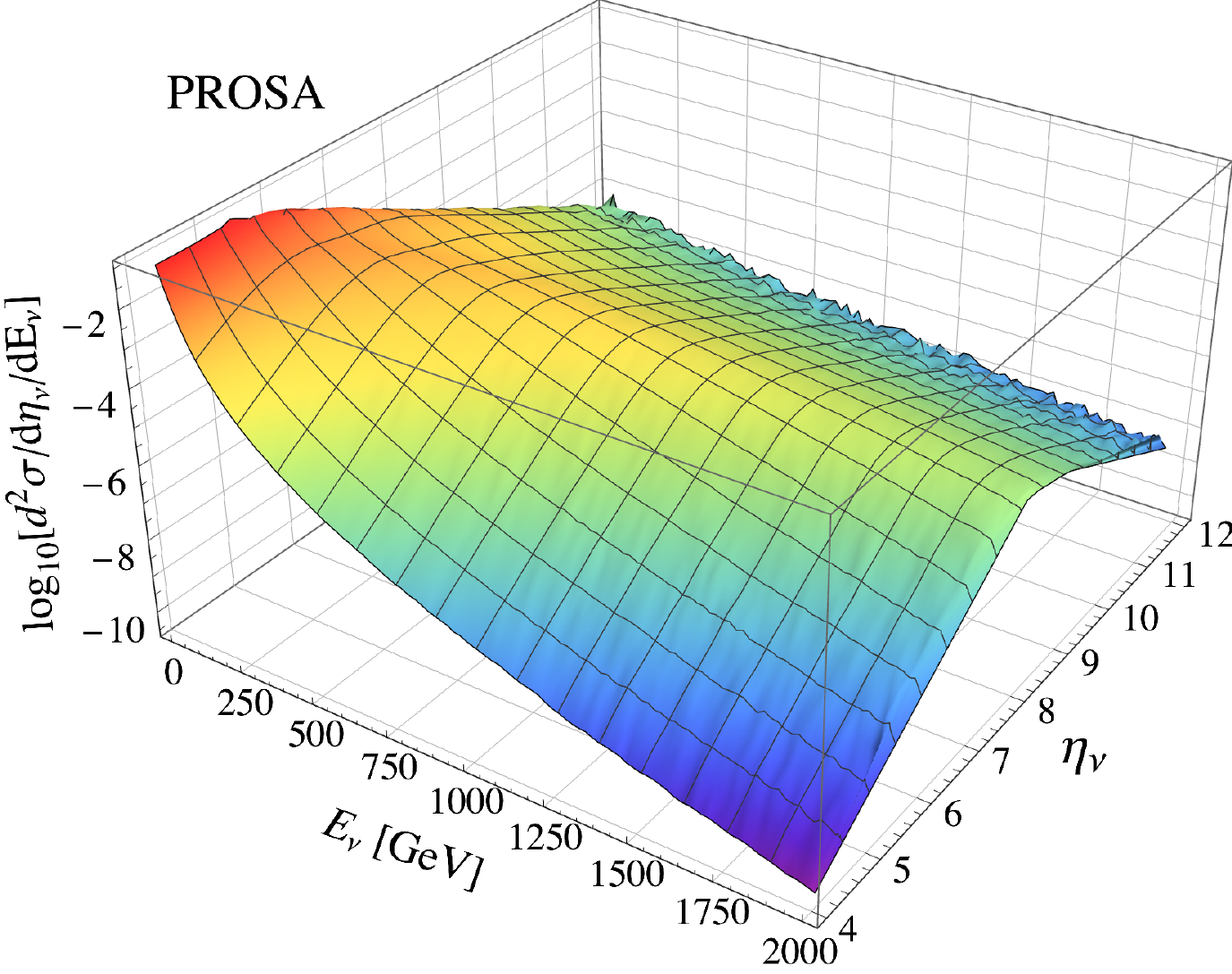}
\includegraphics[width=0.45\textwidth]{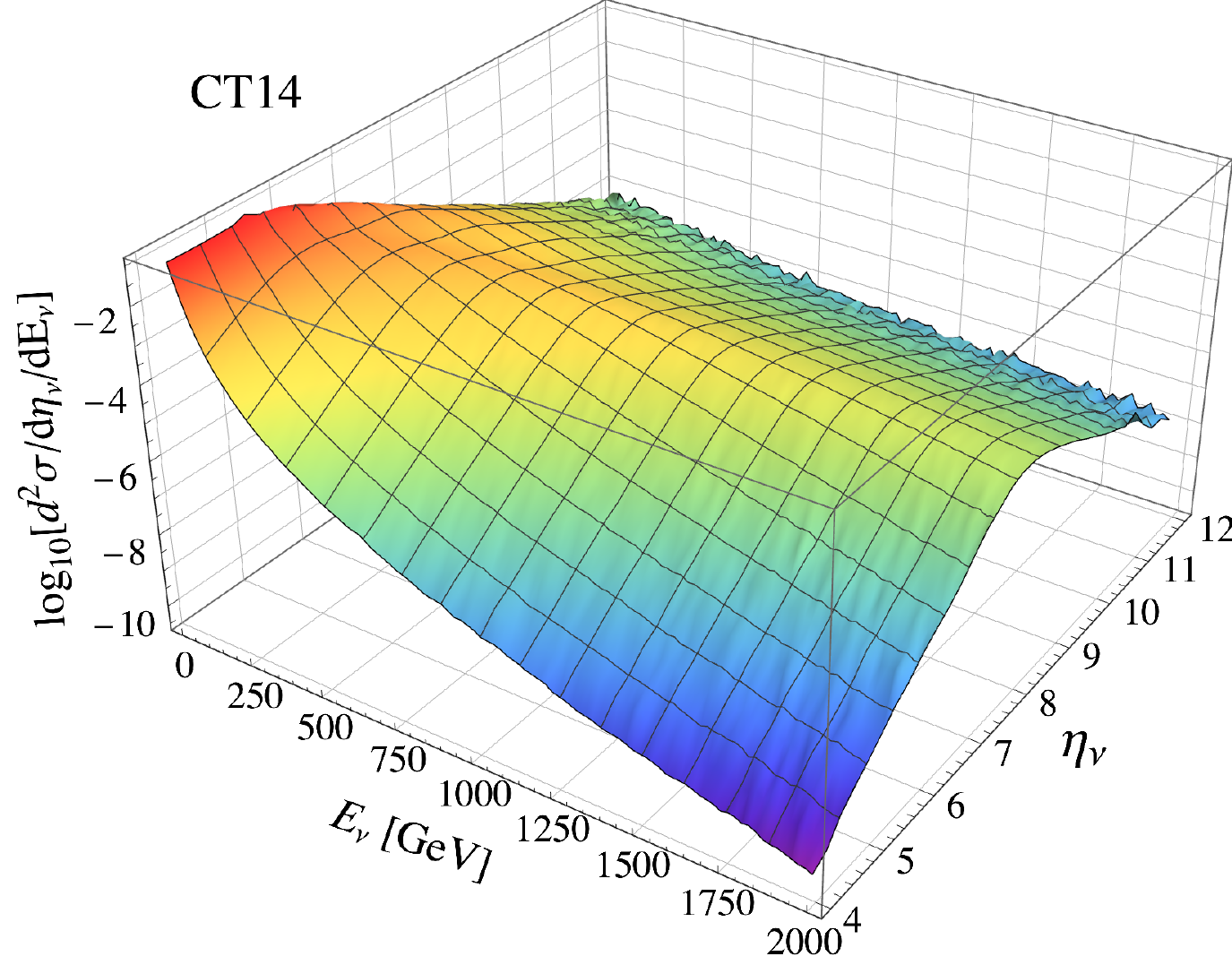}  
\hskip 0.3cm 
\includegraphics[width=0.055\textwidth]{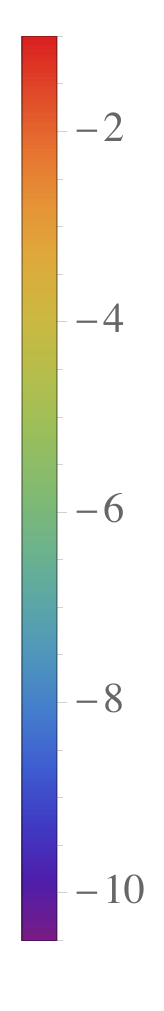}  
    \caption{The distribution $d^2\sigma/(dE_\nu d\eta_\nu)$ in units of $\mu$b/GeV for $\nu_\tau+\bar\nu_\tau$ from $D_s^\pm$ production and decay, evaluated using the PROSA (left) and CT14 (right) PDFs with $\mu_R=\mu_F=m_{T,2}$ and $\langle k_T\rangle=0.7$ GeV.}
    \label{fig:Fnutau3D}
\end{figure}

The PROSA19 PDF \cite{Zenaiev:2019ktw} uncertainties and the QCD scale uncertainties for the $\nu_\tau+\bar\nu_\tau$ energy distributions in the forward region are reported in Ref. \cite{Bai:2021ira} where the neutrino rapidity ranges were restricted to $\eta_\nu>6.9$, $7.2<\eta_\nu<8.6$ and $\eta_\nu > 8.9$. Results for $\eta_\nu>8.5$ are shown in the Snowmass white paper on the Forward Physics Facility \cite{FPFLong:2022tbd}. We find that the dominant uncertainty in the predicted $\nu_\tau+\bar\nu_\tau$ energy distributions comes from the renormalization and factorization scale variation in the NLO evaluation of charm-quark pair production. The uncertainty envelope associated with the dependence on the scales is between a factor of 0.25 to a factor of $\sim 2$ times the distribution with the default scales.
The scale uncertainty translates into the same uncertainty in the number of tau neutrino and antineutrino charged-current events.

Also reported in Ref. \cite{Bai:2021ira} are the $\nu_\tau + \bar{\nu}_\tau$ single-differential
energy distributions for the same rapidity ranges using the 3-flavor NNPDF3.1 \cite{Ball:2017nwa}, CT14 \cite{Dulat:2015mca} and the ABMP16 \cite{Alekhin:2018pai} NLO PDFs, again using as input the default scale choice for the PROSA19 PDFs, namely $\mu_R=\mu_F=m_{T,2}$, and $\langle k_T\rangle = 0.7$ GeV. As noted above, in Ref. \cite{Bai:2021ira}  we also show the NLO QCD predictions using the PROSA19 NLO PDFs with $\mu_F=2\mu_R= 2m_T$ and $\langle k_T\rangle=1.2$ GeV. 
Tables of $\nu_\tau+\bar\nu_\tau$ energy distributions and uncertainties are included in the text of Ref. \cite{Bai:2021ira}, and ascii files of the tables are also available as ancillary files at \url{https://arxiv.org/src/2112.11605v1/anc}.

In the data files associated with this Snowmass white paper, we provide numerical results for double-differential distributions in energy and pseudorapidity of the $\nu_\tau+\bar\nu_\tau$ from $D_s^\pm$ and of the $\nu_e+\bar\nu_e$ from $D^\pm$ for these same scale choices with the PROSA19, NNPDF3.1, CT14 and ABMP16 NLO PDFs. The $\nu_e+\bar\nu_e$ distribution from $D^0$ and $\bar{D}^0$ is very similar to the distribution from $D^\pm$, with differences within a few percent~\cite{Bai:2020ukz}.
The $\nu_e+\bar\nu_e$ are also produced by $D_s^\pm$ and $\Lambda_c^\pm$ decays. The contribution of these channels, however, is only about 13\% and less than 10\% of the contribution from $D^\pm$, respectively \cite{Bai:2020ukz}. 
With these double-differential distributions, event numbers for specific detector shapes and locations on- or off-axis can be evaluated in a straightforward way.
The data tables for the double-differential cross-sections in  neutrino energy and pseudorapidity, using bins of width $\Delta \eta_\nu=0.25$ in the $\eta_\nu=4-12$ range and $\Delta E_\nu=20$ GeV for $E_\nu<2$ TeV, can be found in ancillary files at the arXiv link for this preprint.

Representative results for the double-differential distributions for $\nu_\tau+\bar\nu_\tau$  and $\nu_e+\bar\nu_e$  are shown in the upper and lower panels of Fig.~\cref{fig:3in1}, respectively. They are obtained with the default scales and the PROSA19 NLO PDFs. Each histogram refers to a different range of $\eta_\nu$, from $\eta_\nu=4.0-4.25$ to $\eta_\nu=9.25-9.5$. 

\begin{figure}[t]
    \centering
    \vskip 0.25in
\includegraphics[width=0.45\textwidth]{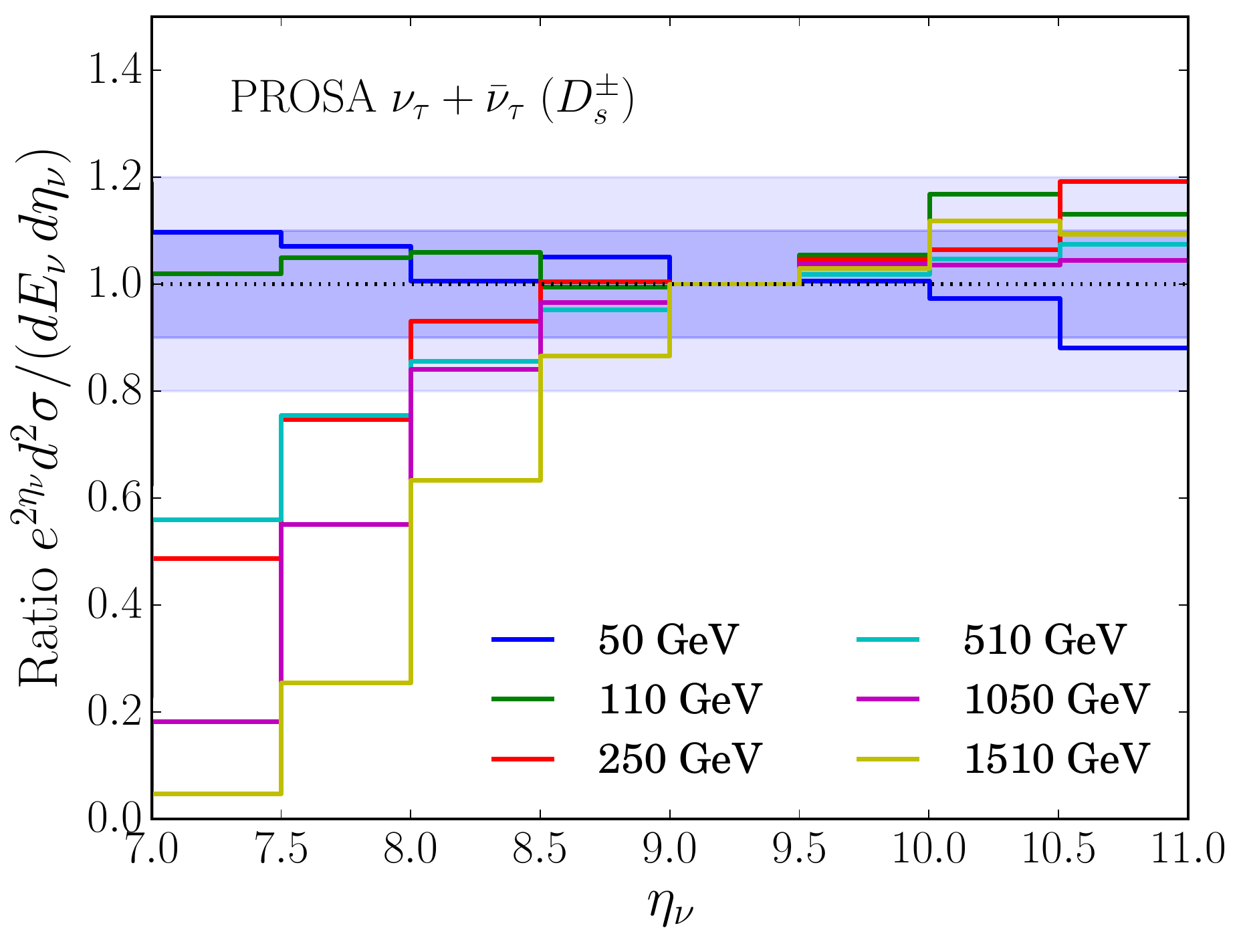}    
\includegraphics[width=0.45\textwidth]{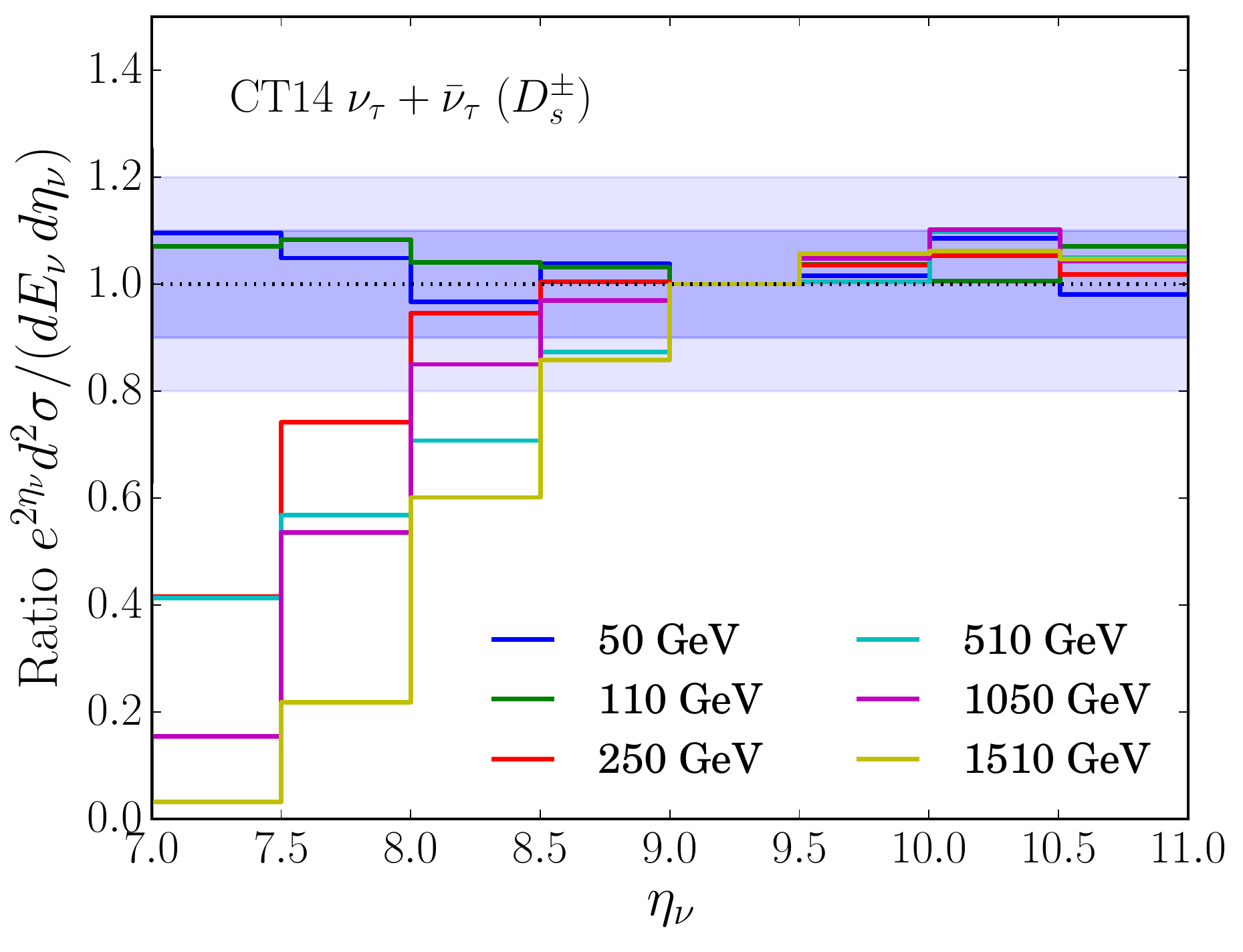}   \includegraphics[width=0.45\textwidth]{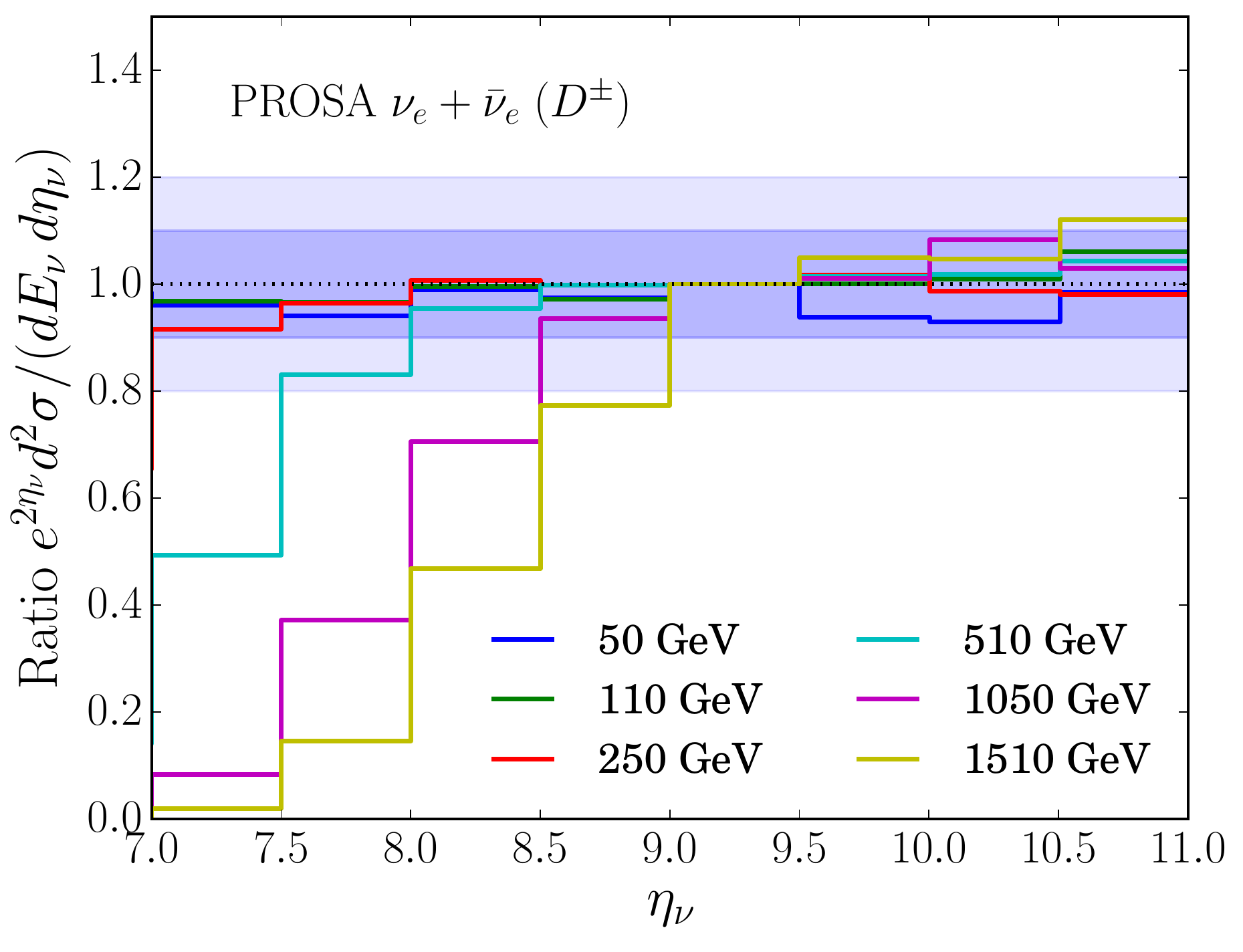}    
\includegraphics[width=0.45\textwidth]{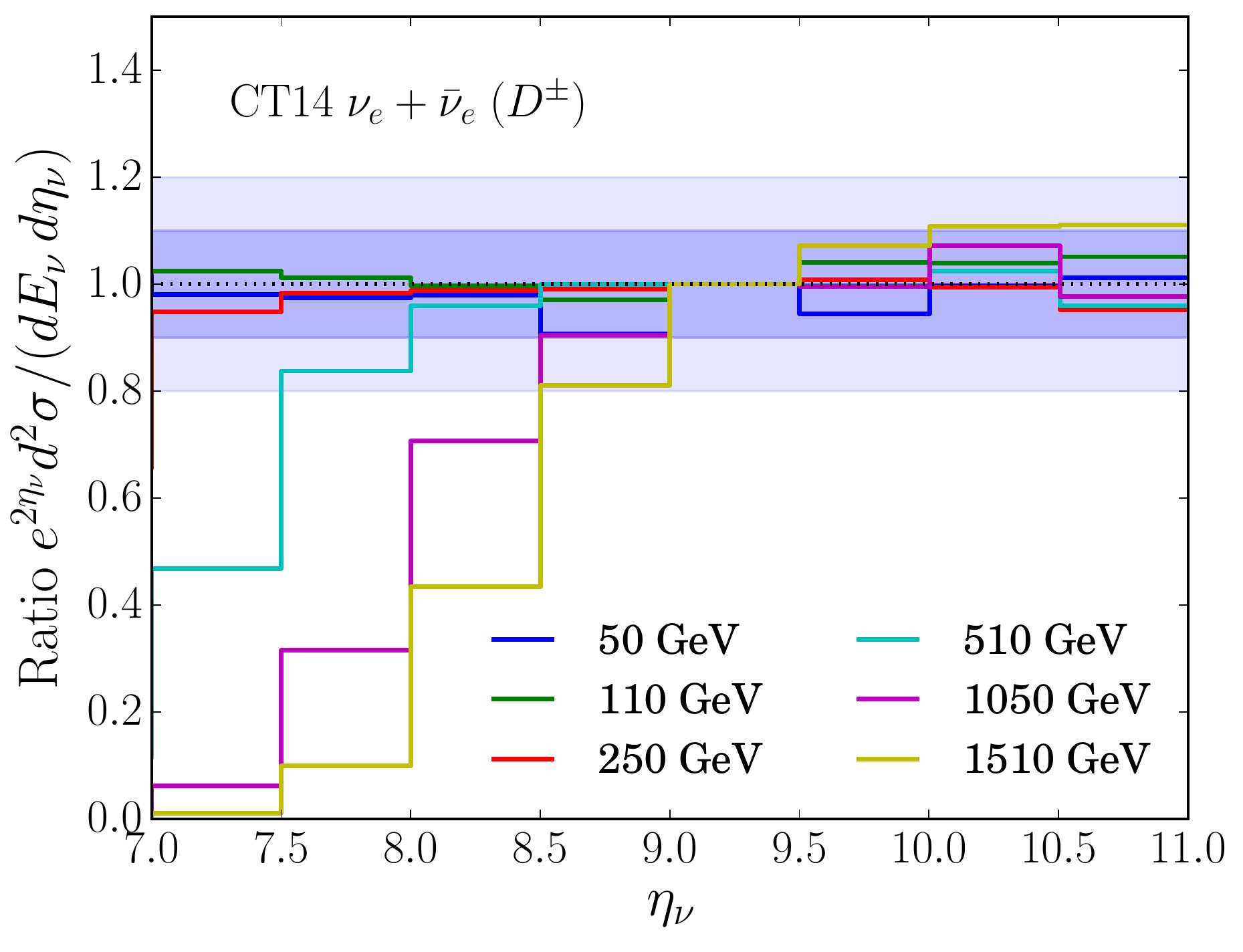}    
    \caption{For $f(E_\nu,\eta_\nu)$ defined in \cref{eq:funf},
    the ratio 
    $f(E_\nu^0,\eta_\nu)/f(E_\nu^0,\eta_\nu^0)$ for selected $E_\nu^0$ with $\eta_\nu^0=9.25$, for PROSA (left) and
    CT14 (right) evaluations of the $\nu_\tau+\bar\nu_\tau$ from $D_s^\pm$ (upper) and $\nu_e+\bar\nu_e$ from $D^\pm$
(lower). The shaded band shows $\pm 10\%$ and $\pm 20\%$ around the ratio equal to 1.}
    \label{fig:etascaling}
\end{figure}

\cref{fig:Fnutau3D} shows the double-differential distribution $d^2\sigma/(dE_\nu d\eta_\nu)$ for $\nu_\tau+\bar\nu_\tau$ for both the PROSA and the CT14 NLO PDF sets. The figure illustrates the feature already observed in the distribution $d\sigma/d\eta_\nu$ for $\eta_\nu\gsim 8.3$, that $d\sigma/d\eta_\nu\sim \exp(-2\eta_\nu)$~\cite{Bai:2021ira,Kling:2021gos}. The double-differential distribution shows that 
similar features manifest
for a wide range of energies and for a larger range of $\eta_\nu$, depending on $E_\nu$. 
For $f(E_\nu,\eta_\nu)$ defined by 
\begin{equation}
\label{eq:funf}
    f (E_\nu,\eta_\nu)\equiv e^{2\eta_\nu}\cdot\frac{d^2\sigma}{dE_\nu\, d\eta_\nu}\,,
\end{equation}
$f(E_\nu,\eta_\nu)$ becomes independent of $\eta_\nu$ for large enough $\eta_\nu$.
\cref{fig:etascaling} shows
$f(E_\nu^0,\eta_\nu)/f(E_\nu^0,\eta_\nu^0)$ for selected $E_\nu^0$ for $\nu_\tau+\bar\nu_\tau$ and $\nu_e+\bar\nu_e$ for two PDF sets. The value of $\eta_\nu^0$ in $f(E_\nu^0,\eta_\nu^0)$ in the denominator of the ratio is for the $\Delta \eta_\nu$ bin centered at $\eta_\nu^0=9.25$.  The dark blue shaded band shows the range of $\eta_\nu$ values where the scaling of the differential cross section as $\sim e^{-2\eta_\nu}$ is valid to within $\pm 10\%$, and the light band shows the scaling to within $\pm 20\%$. 

\section{Future developments}

The large scale uncertainty in the theoretical predictions of  forward neutrino fluxes from heavy flavor assessed in Ref. \cite{Bai:2021ira} is one of several uncertainties. 
Beside the uncertainty related to missing 
 higher-order
perturbative contributions,  further uncertainties are associated to a number of ingredients entering the predictions: parton distribution functions, fragmentation functions, intrinsic transverse momentum, multiple parton interactions and the potential role of intrinsic charm. Some of these ingredients are within the collinear factorization approximation, whereas other ones go beyond it. In particular, quantitative estimates of the role of power corrections will be important, although complex, especially as for predictions in the most forward direction. 

An intermediate step in this program is to study the potential of experimental measurements at LHCb and more forward rapidity measurements to constrain new parton distribution function fits, for fixed fragmentation function inputs, as well as new fragmentation function fits, including a systematic assessment of uncertainties. This theoretical effort, in a different energy regime, is also relevant to make predictions for the DsTau (NA65) experiment \cite{Aoki:2019jry},  designed to detect
$D_s\to\tau\nu_\tau$ with emulsion detectors, where $D_s^\pm$ are produced by a 400 GeV proton beam incident on a tungsten or molybdenum target. Test runs in 2016 and 2017, and a pilot run in 2018 initiated this experiment \cite{Aoki:2019jry,Vasina:2021euh}.

A more complete understanding of forward production of charmed hadrons and their decays to neutrinos \cite{Jeong:2021vqp} might also have some implications for theoretical predictions of the prompt atmospheric neutrino flux \cite{Zenaiev:2019ktw,Bhattacharya:2016jce,Bhattacharya:2015jpa,Benzke:2017yjn,Garzelli:2016xmx}. Hadrons produced by high-energy cosmic ray interactions with air nuclei decay or re-interact in the atmosphere. The very short lifetimes of charmed hadrons relative to pions and kaons allow the prompt neutrinos from charm to dominate the atmospheric neutrino flux at laboratory neutrino energies above $\sim 10^5-10^6$ GeV. Theoretical uncertainties in the prompt neutrino flux impact background evaluations for measurements at the Very Large Volume Neutrino Telescopes, for example by the IceCube Neutrino Observatory\cite{IceCube:2020acn,Stettner:2019tok} or by KM3NeT \cite{KM3net2016}, of the diffuse neutrino flux produced by the highest-energy cosmic accelerators. The kinematical region relevant for the formation of the prompt atmospheric neutrino flux is indeed much larger \cite{Jeong:2021vqp} than the narrow one which will be explored by the Faser$\nu$ and SND@LHC experiments during the Run-III at the LHC. In order to have better chance to constrain charm production in the region relevant for very-large-volume neutrino telescopes applications, it is important to design and place forward experiments sensitive to charm quark production in the rapidity region $4.5 < y < 6$.

\section*{Acknowledgments}

We are grateful to members of the PROSA collaboration for useful discussions on various aspects of the PROSA and other PDF fits of which they are authors. We are grateful to the Snowmass 2021 community for inputs and discussions which helped to shape this research activity.
This work is supported in part by US Department of Energy grants DE-SC-0010113, DE-SC-0012704, by  German BMBF contract 05H21GUCCA, 
and by the National Research Foundation of Korea (NRF) grant funded by the Korea government through Ministry of Science and ICT grant number No. 2021R1A2C1009296.

\bibliographystyle{JHEP}

\providecommand{\href}[2]{#2}\begingroup\raggedright\endgroup

\end{document}